# Single Channel Cutaneous Electrogastrography Parameters from Local White Rabbit (*Oryctolagus cuniculus*)

## Parameter Elektrogastrografi Kutan Satu *Channel* dari Kelinci Putih Lokal (*Oryctolagus cuniculus*)

Tyas Pandu Fiantoro[1], Lussya Eveline Rawar[2]

[1]Electronic Engineering undergraduate, Universitas Gadjah Mada
[2]Veterinary Sciences undergraduate, Universitas Gadjah Mada
Email: eltroyaz@mail.ugm.ac.id

**Abstract**

EGG recordings performed on 13 local white rabbits (*O. cuniculus*) which divided into 3 groups; acetosal 35 mg/kgBM receiver, reserpine 37.5 mg/kgBM receiver, and control group. A total of 72 EGG recordings obtained from them, which divided furthermore into nine datasets based on prepandrial state, postpandrial state, and post 1 hour drug administration state. EGG parameters such as the number of cycle per minute (*cpm*), average voltage of action potential segment ($\bar{V}_a$), root mean square voltage of action potential segment ($A_{rms}$), root mean square voltage of all EGG segment ($V_{rms}$), average period of action potential segment ($\bar{T}_a$), average period of resting plateau ($\bar{T}_l$), average period difference among action potential segment and resting plateau ($\bar{T}_a - \bar{T}_l$), and dominant frequency ($f_d$) are obtained. Insignificant difference of $f_d$ ($P$ = 0.9112993) and *cpm* ($P$ = 0.9382463) within nine EGG datasets was found. These findings contrasted the common practice of EGG assessment, which $f_d$ and *cpm* are the main parameters for diagnosis base. In other hand, significant difference between nine EGG datasets found for $\bar{V}_a$, $A_{rms}$, and $V_{rms}$ parameter with $P$ = 0.0007346, 0.0039191, and 0.0000559 respectively. In conclusion, EGG parameterization should not be limited to $f_d$ and *cpm* only.

**Abstrak**

Perekaman EGG dilakukan pada 13 ekor kelinci putih lokal (*O. cuniculus*) yang dibagi menjadi 3 kelompok perlakuan, yakni kelompok yang diberikan asetosal PO 37,5 mg/kgBB, reserpin PO 5 mg/kgBB, dan kontrol. Diperoleh 72 rekaman EGG dari tiga kelompok perlakuan kelinci tersebut, yang kemudian dikelompokkan menjadi 9 set data berdasarkan keadaan puasa (*prepandrial*), setelah makan (*postpandrial*), dan satu jam paska pemberian asetosal atau reserpin. Parameter-parameter EGG berupa *cycle per minute* (*cpm*), rerata tegangan segmen potensial aksi EGG ($\bar{V}_a$), *root mean square* tegangan segmen potensial aksi EGG ($A_{rms}$), *root mean square* semua segmen EGG ($V_{rms}$), rerata durasi potensial aksi ($\bar{T}_a$), rerata durasi *plateau* istirahat ($\bar{T}_l$), rerata selisih dari durasi potensial aksi dengan durasi *plateau* istirahatnya ($\bar{T}_a - \bar{T}_l$), dan frekuensi dominan ($f_d$) diperoleh dari hasil pengolahan data rekaman EGG. Ditemukan hasil bahwa $f_d$ antar 9 set data EGG ialah tidak berbeda signifikan ($P$ = 0,9112993), demikian juga dengan parameter *cpm* ($P$ = 0,9382463). Hal ini kontras dengan praktek umum EGG yang menjadikan $f_d$ dan *cpm* sebagai pertimbangan diagnosa. Akan tetapi, ada perbedaan signifikan pada suatu golongan kelinci (*O. cuniculus*) untuk parameter $\bar{V}_a$ ($P$ = 0,0007346), $A_{rms}$ ($P$ = 0,0039191), dan $V_{rms}$ ($P$ = 0,0000559). Dengan demikian, parameterisasi EGG disarankan tidak hanya terpaku pada $f_d$ dan *cpm*.

**Kata kunci:** *Oryctolagus cuniculus*, elektrogastrografi, parameter, nilai *P*, lambung

## Introduction

Cutaneous electrogastrography (commonly abbreviated as EGG) is a method to record gastric electrical activity cutaneously. This technique promises a non-invasive gastric assessment method that even can be made as a mobile diagnostic device.

This article focused on the EGG acquisition parameters and hypothesis testing based on Fischer's one tail test. The hypothesis testing performed on each EGG parameter, to reveal its differentiability for nine EGG datasets. If there is any parameter that could be used to differentiate each dataset, then further design for a non-invasive monitoring of the gastric fed state could be performed easily. This promises many benefits such as gastric early warning system (EWS), non-invasive nutrition-uptake sensor, and even play a role in gastric disease prevention such as gastritis and gastroesophageal reflux disease (GERD).

## Material and Methods

Thirteen local white rabbits (*O. cuniculus*) divided into three groups; each rabbit in the first group receives acetosal orally, each rabbit in the second group group receives reserpine orally, and the the third group as the control group. Acetosal was given for triggering gastric mucosal epithelium erosion (Konturek *et al.*, 1982), whereas reserpine was given for triggering excessive gastric acid secretion (Schneider & Clark, 1956; Greenbaum, *et al*, 1959). All groups fed between 14.00 and 16.00, with *ad libitum* supply of drinking water given all day. All rabbits in the first group were given an oral suspension of acetosal with a dose of 37,5 mg/kgBM once a day, and all rabbits in the second group were given an oral suspension of reserpine with a dose of 5 mg/kgBM once a day. Both drug administration performed at 16.00 when all the accessible foods in the cage were cleared out. All rabbits were fasted between 16.00 and the 14.00 of the next day, drinking water still available. Oral drug administration performed using an oral syringe. All treatments (and recordings) last for three days. A 2 mm hosed bottle of 300 ml capacity was used as each rabbit's drinking bottle, and refilled with tap water twice a day. Fodder used was one type of grass called the polar grass.

EGG recording accomplished by attaching a pair of cutaneous electrode with diameter of 1.5 mm (*PRYM Press Stud 555*) on the already shaved (using WAHL clipper) lumbar surface. The shaved lumbar area has a fur thickness around 2 mm. The two electrodes consisted of a red electrode and a green electrode. The red electrode placed on the lumbar area that represents a point between *esophageal sphincter* and *orad corpus*, and the green electrode placed on the lumbar area that represents a point between *pyloric sphincter* and *terminal anthrum* as shown in Figure 1 (Kenneth dkk., 2004; Sagami dkk., 2007; Květina dkk., 2010; Tyas dan Eveline, 2014). Each of the electrodes connected into the input pins of AD-620 (*Texas Instrument*) instrumentation amplifier that configured to produce a gain value of 50. The output of the instrumentation amplifier fed into a digital oscilloscope (GWINSTEK GDS 1102-A-U). The final result is a .csv file that exported into a flash disk (*HP v175w 16 GB*) that attached to the oscilloscope beforehand. Before the electrodes were attached, a conductive gel was given on the respective lumbar area. To make sure the attached electrodes did not detach, a bandage (Hansaplast) of 2 cm width and 3 cm to 4 cm length is used. This method did not hurt the experiential rabbits (*O. cuniculus*).

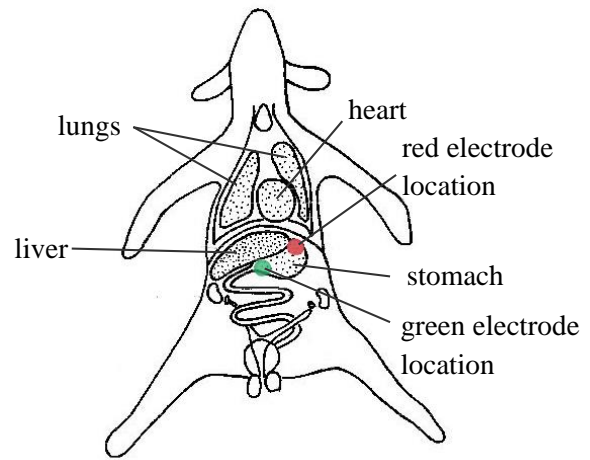

Figure 1. Illustration of the electrode placement

EGG collected twice a day for each individual for 3 days of observation. The recording was done before the individual fed

(between 12.00 and 14.00), and after the individual had been fed and/or given drug (between 16.00 and 18.00). The recording performed in 400 seconds with a sampling rate of 10 Hz. This considered sufficient (Jieyun Yin, 2013) because *Oryctolagus cuniculus* observed to be having gastric wave rate 3 to 4 times faster compared to *Homo sapiens*.

### EGG Datasets

A total of 72 EGG recordings were obtained. These recordings then divided into nine datasets based on the fed condition (prepandrial – postpandrial), and the drug given (acetosal, reserpine, or given nothing). The nine datasets are described as the following

1) A population of prepandrial individuals
2) A population of postpandrial individuals
3) Prepandrial control individuals
4) Postpandrial control individuals
5) Prepandrial acetosal recipients
6) Postpandrial acetosal recipients
7) All reserpine recipients
8) All control individuals
9) All acetosal recipients

### EGG Parameterization

Obtained EGG then processed to extract its parameters. The EGG parameters examined are cycle per minute (*cpm*), average voltage of action potential segments ($\bar{V}_a$), root mean square voltage of action potential segments ($A_{rms}$), root mean square voltage of all EGG segments ($V_{rms}$), average duration of action potential segments ($\overline{T_a}$), average duration of resting potential segments ($\overline{T_l}$), the average duration difference between each corresponding action potential segment and resting potential segment ($\overline{T_a - T_l}$), and dominant frequency ($f_d$).

The parameters mentioned are assessed using these methods; *cpm* is the result of dividing total cycle number with the recording's duration in minutes (so the name "cycle per minute"), $V_{rms}$ (in milivolt) obtained from taking the root mean square of all samples of one EGG recording, $A_{rms}$ (in milivolt) obtained from taking the root mean square of all samples but resting potential segment of one EGG recording, $\bar{V}$ (in milivolt) acquired by taking the average of all samples of one EGG recording EGG, $\overline{T_a}$ (in seconds) yielded from dividing the duration sum of depolarisation – plateau – repolarisation segments with the total cycle number from one recording, $\overline{T_l}$ (in seconds) produced from dividing the duration sum of resting segments with the total cycle number from one recording, $\overline{T_a - T_l}$ (in seconds) obtained by averaging the difference among each corresponding action potential duration and resting duration, and $f_d$ (in milihertz) obtained from taking the Fast Fourier Transform (FFT) for one recording and taking the frequency with the highest spectral power (with the exception for frequencies that below 6 mHz because this considered as the steady or DC component of the recording).

### Results

The statistics of eight EGG parameters (*cpm*, $\bar{V}_a$, $A_{rms}$, $V_{rms}$, $\overline{T_a}$, $\overline{T_l}$, $\overline{T_a - T_l}$, dan $f_d$) for nine datasets given in Figure 2 and 3. *P* value calculated with one tailed Fischer's test for respective parameter in nine datasets. The test was accomplished using SCILAB 5.5.1. For all parameters those *P* value lower than 0.01, would be considered for further differential observation such as between prepandrial and potpandrial datasets, and among drug recipient dataset, reserpine recipient dataset, and control dataset. This further observation's goal is to determine which parameter that could be used for differentiating fed states and/or drug recipients.

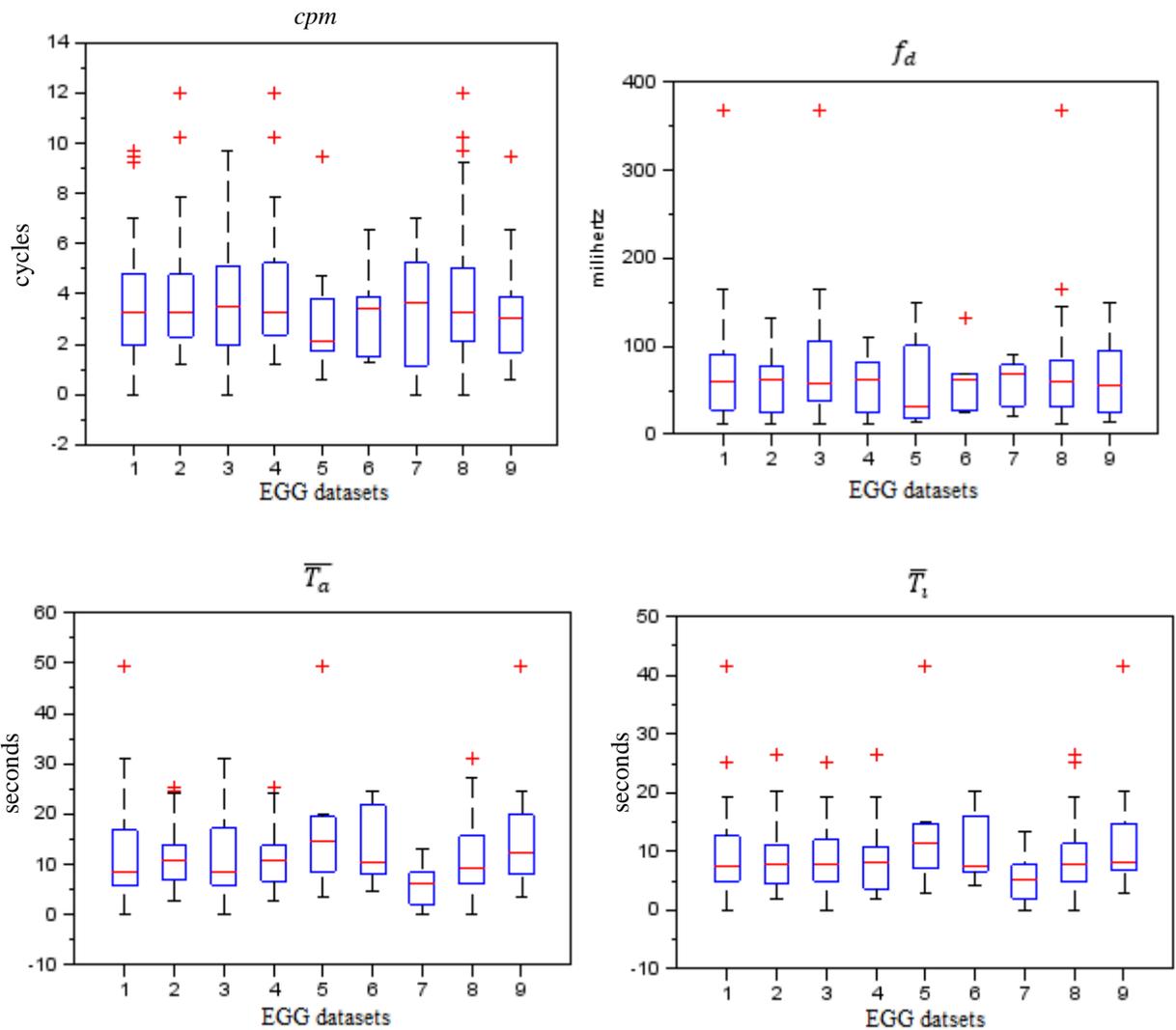

Figure 2. Each $f_d$, $cpm$, $\overline{T_a}$ and $\overline{T_\iota}$ parameter statistics for all datasets.

It can be seen in Figure 2 that each $f_d$, $cpm$, $\overline{T_a}$ and $\overline{T_\iota}$ parameter has a–hard–to–distinguish top quartile, bottom quartile, median, and range within nine datasets, because all of the parameters mentioned has $P > 0.2$. This indicate that the differences among them may be caused from a mere data randomness, moreover for the ones with $P > 0.9$.

The common EGG parameters assessed are $f_d$ and $cpm$, that already becoming the base of tachygastria, normogastria, or bradygastria gastric condition diagnosis (Kentie *et al.*, 1981; Rebrov *et al.*, 1996; Vornovitsky dan Feldshtein, 1998; Parkman *et al.*, 2003; Chang, 2005; Tropskaya *et al.*, 2005; Glinina and Shkatova, 2007; Pasechniko *et al.*, 2007). In this research, both $f_d$ and $cpm$ observed to lack of significant difference within the nine datasets ($f_d$ has $P = 0.9112993$, and $cpm$ has $P = 0.9382463$).

The $P$ values for $\overline{T_a}$ and $\overline{T_\iota}$ within nine datasets are 0.2238246 and 0.3428104 respectively. It is shown in Figure 2 that both $\overline{T_a}$ and $\overline{T_\iota}$ parameters have the similar ranges and quartiles. Actually, as observed in Figure 2, all the four parameters are hard to distinguish within the nine datasets. Therefore, these parameters ($f_d$, $cpm$, $\overline{T_a}$, and $\overline{T_\iota}$) are not recommended for differentiating the used datasets.

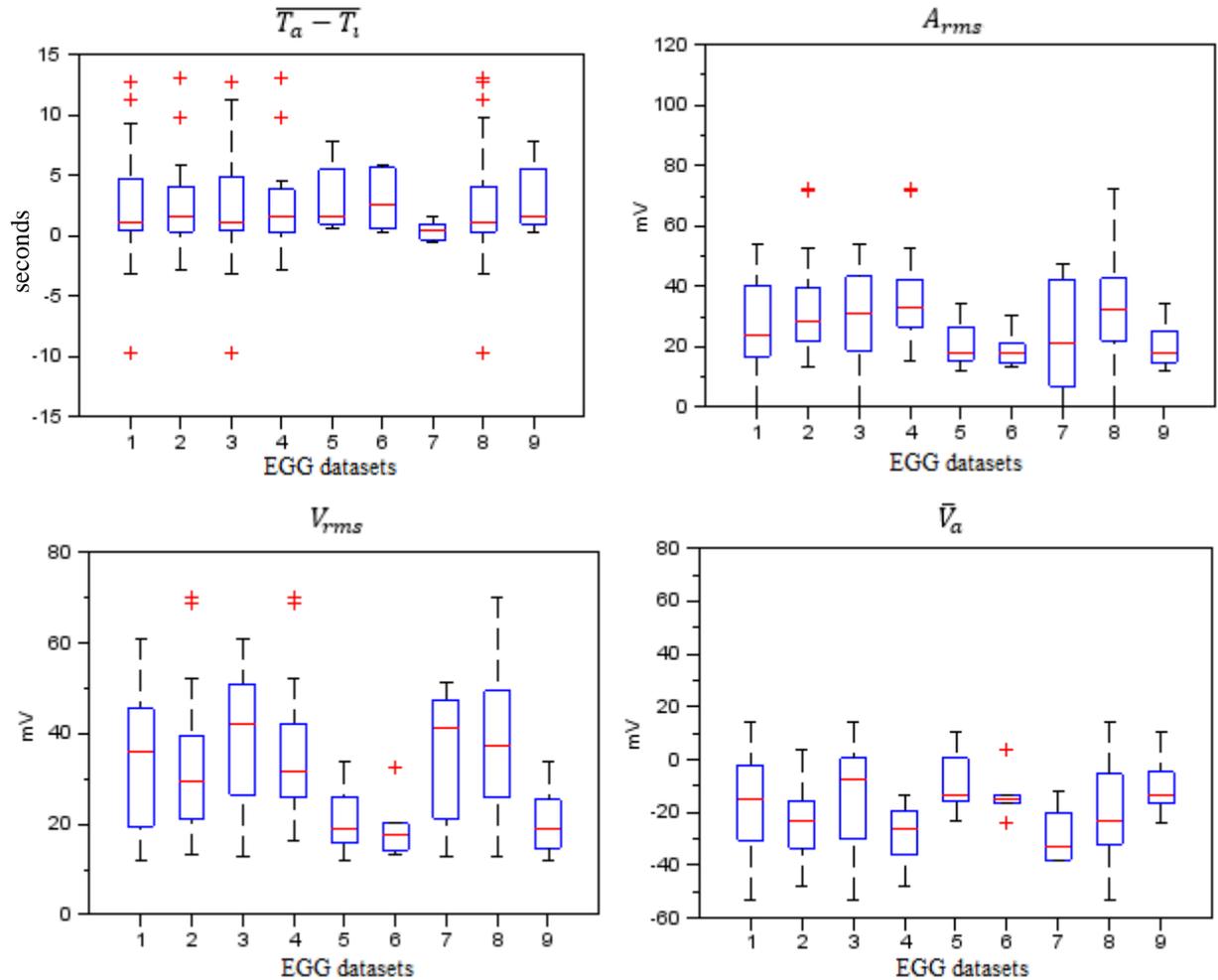

Figure 3. Each $\overline{T_a - T_\iota}$, $A_{rms}$, $V_{rms}$, and $\bar{V}_a$ parameter statistics for all datasets.

Even from previous observation suggests that both for $\overline{T_a}$ and $\overline{T_\iota}$ has a P value larger than 0.2, the P value of the difference of the duration of action potential segments with the duration of the accompanying resting potential segment ($\overline{T_a - T_\iota}$) need to be confirmed. This differential parameter is measured in seconds. Positive $\overline{T_a - T_\iota}$ value means averagely the action potential segments collective duration is longer than its accompanying resting potential segment duration, whereas a negative value of $\overline{T_a - T_\iota}$ means averagely the action potential segments collective duration is shorter than its accompanying resting potential segment duration.

A fact of $P = 0.9486937$ obtained for the $\overline{T_a - T_\iota}$ parameter, as it can be seen in Figure 3 that this parameter does not have a distinguishable feature within the nine datasets. Its quartiles, ranges, and medians are indistinctive within the datasets, an exception for the 7[th] dataset. This 7[th] dataset belongs to the reserpine 5 mg/kgBM receiver group. Every rabbit (*O. cuniculus*) in this group seen to be disphagic, because of the hipotensic effect of the reserpine given. Hence, it is hard to differentiate which fed state the individual had. This backs up the reason of the merging of the prepandrial state with the postpandrial state as one grouping. Euthanasia follows all rabbit (*O. cuniculus*) that observed disphagic shortly. This was done to prevent unprecedented death, so the gastric physical parameters could be accessed.

It could be observed in Figure 3, that there is a distinguishable feature for each $A_{rms}$, $V_{rms}$, and $\bar{V}_a$ parameter within the nine datasets. The prominent feature for those parameters are; all of those parameters ($A_{rms}$, $V_{rms}$, and $\bar{V}_a$) had the quartiles and medians that always higher for dataset 1 and 3 (prepandrial fed state datasets) compared with dataset 2 and 4 (postpandrial fed state datasets), the $A_{rms}$ and $V_{rms}$ parameters were also always had the quartiles and domains that higher for dataset 8 (all control individuals) compared to the 9th dataset (acetosal receivers), only the $\bar{V}_a$ parameter were observed as having the quartiles and domains that lower for dataset 8 (all control individuals) compared to the 9th dataset (acetosal receivers).

Each of the $A_{rms}$, $V_{rms}$, and $\bar{V}_a$ parameter, had a *P* value that less than *0.01* (respectively 0.0007346, 0.0039191, and 0.0000559) for the nine datasets domain. From this fact, a further one tailed Fischer's test was performed in stricter domains. The new testing domains were determined based on the fed state variability and the administered drug consideration. These new domains just limited into two datasets.

Table 1. *P* values for $A_{rms}$, $V_{rms}$, and $\bar{V}_a$ parameters, within stricter domains.

| *P* | Dataset Domain | | | | | | | |
|---|---|---|---|---|---|---|---|---|
| | 1 and 2 | 3 and 4 | 5 and 6 | 3 and 5 | 4 and 6 | 7 and 8 | 7 and 9 | 8 and 9 |
| $A_{rms}$ | 0.218619 | 0.257351 | 0.61103 | 0.059126 | 0.014089 | 0.109347 | 0.543076 | 0.001985 |
| $V_{rms}$ | 0.5743211 | 0.593029 | 0.609566 | 0.001732 | 0.011138 | 0.717166 | 0.002929 | 0.000042 |
| $\bar{V}_a$ | 0.0212452 | 0.004 | 0.470095 | 0.442547 | 0.006546 | 0.153538 | 0.000818 | 0.051489 |

The *P* values for $A_{rms}$, $V_{rms}$, and $\bar{V}_a$ parameters within certain two datasets presented in Table 1. There are prominent features for $\bar{V}_a$ parameter, those it could be used to distinguish prepandrial control individuals from the postpandrial control individuals (3rd dataset from the 4th dataset), postpandrial control individuals from the postpandrial acetosal receivers (4th dataset from the 6th dataset), and reserpine receivers from acetosal receivers (7th dataset and the 9th dataset). Similar features also observed for $V_{rms}$ parameter, those it could be used to distinguish prepandrial control individuals from the postpandrial control individuals (3rd dataset from the 4th dataset), prepandrial control individuals from the prepandrial acetosal receivers (3rd dataset from the 5th dataset), reserpine receivers from acetosal receivers (7th dataset and the 9th dataset), and control individuals from acetosal receivers (8th dataset and the 9th dataset). The $A_{rms}$ is only able to distinguish control individuals from acetosal receivers (8th dataset and the 9th dataset). All of these statements supported by the *P* values those less than 0.003.

Thus, these parameters ($A_{rms}$, $V_{rms}$, and $\bar{V}_a$) could be used for EGG dataset differentiation between the mentioned domains. As all of these three parameters are based on amplitude information, this finding is in accordance with the Russian EGG researches (Rebrov, 1981; Biryaltsev *et al.*, 2003; Zakirov *et al.* 2005; Smirnova *et al.* 2009).

**Conclusion**

Statistical evidence showed that EGG parameters such as $f_d$, *cpm*, $\overline{T_a}$, $\overline{T_i}$, and $\overline{T_a - T_i}$ may not be significant (all have $P > 0.2$) for assessing gastric fed state and post drug (acetosal or reserpine) administration gastric state. This finding suggest that tachygastric, normogastric, dan bradygastric condition may not be relevant for differentiating gastric fed within the used nine datasets.

Therefore, EGG parameterization should not be limited to $f_d$ and *cpm* only. The segmental EGG amplitude parameters ($A_{rms}$, $V_{rms}$, and $\bar{V}_a$) may have a substantial usage for diagnosing gastric fed state, and post drug administration gastric state.

**Discussion**

Thirteen rabbits (*O. cuniculus*) were used in this research, but only 72 EGG recordings obtained from them. It is highly recommended to replicate this research with more EGG recordings, thus minimizing the risk of artifacts such as sampling error.

Every treatment given for each individual must be monitored effectively. Hence, for further research, it also recommended that blood plasma drug concentration to be monitored, along with gastric mucosal epithelium measurement together with physical gastric content parameter assessment. The monitoring should be performed in real time as opposed to this research, where the gastric physical measurement only performed after euthanasia due to technical limitation.